\documentclass[twocolumn]{aastex6}
\usepackage{multirow}
\accepted{July 17, 2017}

\shorttitle{Blind mm line emitter search using ALMA data toward gravitational lensing clusters}
\shortauthors{Yamaguchi et al.}

\begin{document}

\title{Blind millimeter line emitter search using ALMA data toward gravitational lensing clusters}

\email{yyamaguchi@ioa.s.u-tokyo.ac.jp}

\author{Yuki Yamaguchi\altaffilmark{1}}

\author{Kotaro Kohno\altaffilmark{1,2}}

\author{Yoichi Tamura\altaffilmark{1,3}}

\author{Masamune Oguri\altaffilmark{4,2,5}}

\author{Hajime Ezawa\altaffilmark{6,7}}

\author{Natsuki H. Hayatsu\altaffilmark{5}}

\author{Tetsu Kitayama\altaffilmark{8}}

\author{Yuichi Matsuda\altaffilmark{9,7}}

\author{Hiroshi Matsuo\altaffilmark{10,7}}

\author{Tai Oshima\altaffilmark{11,7,10}}

\author{Naomi Ota\altaffilmark{12}}

\author{Takuma Izumi\altaffilmark{9}}

\author{Hideki Umehata\altaffilmark{13,1}}

\altaffiltext{1}{Institute of Astronomy, Graduate School of Science, The University of Tokyo, 2-21-1 Osawa, Mitaka, Tokyo 181-0015, Japan}
\altaffiltext{2}{Research Center for the Early Universe, Graduate School of Science, The University of Tokyo, 7-3-1, Hongo, Bunkyo, Tokyo 113-0033, Japan}
\altaffiltext{3}{Division of Particle and Astrophysical Science, Nagoya University, Furocho, Chikusa, Nagoya 464-8602, Japan}
\altaffiltext{4}{Kavli Institute for the Physics and Mathematics of the Universe (Kavli IPMU, WPI), The University of Tokyo, Chiba 277-8583, Japan}
\altaffiltext{5}{Department of Physics, Graduate School of Science, The University of Tokyo, 7-3-1 Hongo, Bunkyo-ku, Tokyo 113-0033, Japan}
\altaffiltext{6}{Chile Observatory, National Astronomical Observatory of Japan (NAOJ), National Institutes of Natural Sciences (NINS), 2-21-1, Osawa, Mitaka, Tokyo 181-8588, Japan}
\altaffiltext{7}{Department of Astronomical Science, School of Physical Science, SOKENDAI (The Graduate University for Advanced Studies), 2-21-1, Osawa, Mitaka, Tokyo 181-8588, Japan}
\altaffiltext{8}{Department of Physics, Toho University, 2-2-1 Miyama, Funabashi, Chiba 274-8510, Japan}
\altaffiltext{9}{National Astronomical Observatory of Japan (NAOJ), National Institutes of Natural Sciences (NINS), 2-21-1 Osawa, Mitaka, Tokyo 181-8588, Japan}
\altaffiltext{10}{Advanced Technology Center, National Astronomical Observatory of Japan (NAOJ), National Institutes of Natural Sciences (NINS), 2-21-1, Osawa, Mitaka, Tokyo 181-8588, Japan}
\altaffiltext{11}{Nobeyama Radio Observatory, National Astronomical Observatory of Japan (NAOJ), National Institutes of Natural Sciences (NINS), 462-2, Nobeyama, Minamimaki, Minamisaku, Nagano 384-1305, Japan}
\altaffiltext{12}{Department of Physics, Nara Women's University, Kitauoyanishi-machi, Nara, Nara 630-8506, Japan}
\altaffiltext{13}{The Open University of Japan, 2-11 Wakaba, Mihama-ku, Chiba 261-8586, Japan}



\begin{abstract}
We present the results of a blind millimeter line emitter search using ALMA Band 6 data with a single frequency tuning toward four gravitational lensing clusters (RXJ1347.5$-$1145, Abell S0592, MACS J0416.1$-$2403, and Abell 2744). We construct three-dimensional signal-to-noise ratio (S/N) cubes with 60 MHz and 100 MHz binning, and search for millimeter line emitters. We do not detect any line emitters with a peak S/N $>$ 5, although we do find a line emitter candidate with a peak S/N $\simeq$ 4.5. These results provide upper limits to the CO(3-2), CO(4-3), CO(5-4), and [CII] luminosity functions at $z\simeq$ 0.3, 0.7, 1.2, and 6, respectively. Because of the magnification effect of gravitational lensing clusters, the new data provide the first constraints on the CO and [CII] luminosity functions at unprecedentedly low luminosity levels, i.e., down to $\lesssim$ 10$^{-3}$--10$^{-1}$ Mpc$^{-3}$ dex$^{-1}$ at $L^\prime_\mathrm{CO}\sim$ 10$^8$--10$^{10}$ K km s$^{-1}$ pc$^2$ and $\lesssim$ 10$^{-3}$--10$^{-2}$ Mpc$^{-3}$ dex$^{-1}$ at $L_\mathrm{[CII]}\sim$ 10$^8$--10$^{10}$ $L_\odot$, respectively. Although the constraints to date are not stringent yet, we find that the evolution of the CO and [CII] luminosity functions are broadly consistent with the predictions of semi-analytical models. This study demonstrates that the wide observations with a single frequency tuning toward gravitational lensing clusters are promising for constraining the CO and [CII] luminosity functions.
\end{abstract}




\section{Introduction} \label{sec:intro}
Recent studies have unveiled the cosmic star formation history based on multi-wavelength observations \citep[e.g.,][and reference therein]{madau2014, bouwens2015}. The cosmic star formation rate density (SFRD) has a peak level between $z\sim3$ and $z\sim1$, and it subsequently decreases rapidly towards $z=0$. However, the role of dust-obscured star formation at high redshifts (especially at $z\gtrsim$ 3--4) and the physical cause governing the cosmic star formation history are still uncertain.

One of the promising ways to resolve these questions is to observe (sub-)millimeter emission lines. The [CII] 158 $\mu$m line is expected to be a tracer of dust-obscured star formation in local to distant galaxies \citep[e.g.,][]{delooze2011, delooze2014, smail2011, sargsyan2012, sargsyan2014}. The molecular gas content of galaxies can be observed via CO rotational transition lines \citep[e.g.,][]{solomon1987, tacconi2013}. The molecular gas mass of galaxies is one of the fundamental properties to understand the cause of cosmic star formation history because the molecular phase of the interstellar medium is considered as the fuel for star formation activities. However, observations of (sub-)millimeter emission lines have been limited to follow-up studies of galaxies, which are preselected by optical, near-infrared (NIR), or (sub-)millimeter wavelengths \citep[e.g.,][and references therein]{daddi2010, tacconi2010, tacconi2013, tacconi2017, carilli2013, genzel2015}. In these cases, the selection is based on the star formation properties or stellar mass of a given galaxy. Accordingly, these samples are biased.

Based on the above reasons, constraining the luminosity functions of (sub-)millimeter line emitters via unbiased (sub-)millimeter line emitter surveys is necessary to unveil the cosmic star formation history. For example, the ``line intensity mapping'' technique is one of the useful ways to constrain luminosity functions \citep[e.g.,][]{keating2016}. So far, individual properties of line emitters have remained unexplored because the emission from a multitude of galaxies over a wide range of line luminosities are integrated in this ``line intensity mapping'' technique. 

Because of the development of observational instruments such as the IRAM Plateau de Bure Interferometer (PdBI), or NOEMA, and the Atacama Large Millimeter/submillimeter Array (ALMA), unbiased (sub-)millimeter line emitter searches are now feasible \citep[e.g.,][]{walter2014, walter2016, decarli2014, decarli2016, aravena2016}. However, such line emitter searches based on spectroscopic scan observations (i.e., observed frequency range $>$ several tens of gigahertz) can often be expensive in terms of total observing time. Therefore, serendipitous detections of line emitters \citep[e.g.,][]{tamura2014, umehata2017, hayatsu2017} and line emitter searches using archival data \citep[e.g.,][]{matsuda2015, miller2016} based on high-sensitivity observations of ALMA have been reported.

\begin{figure*}[!]
\epsscale{1.}
\plotone{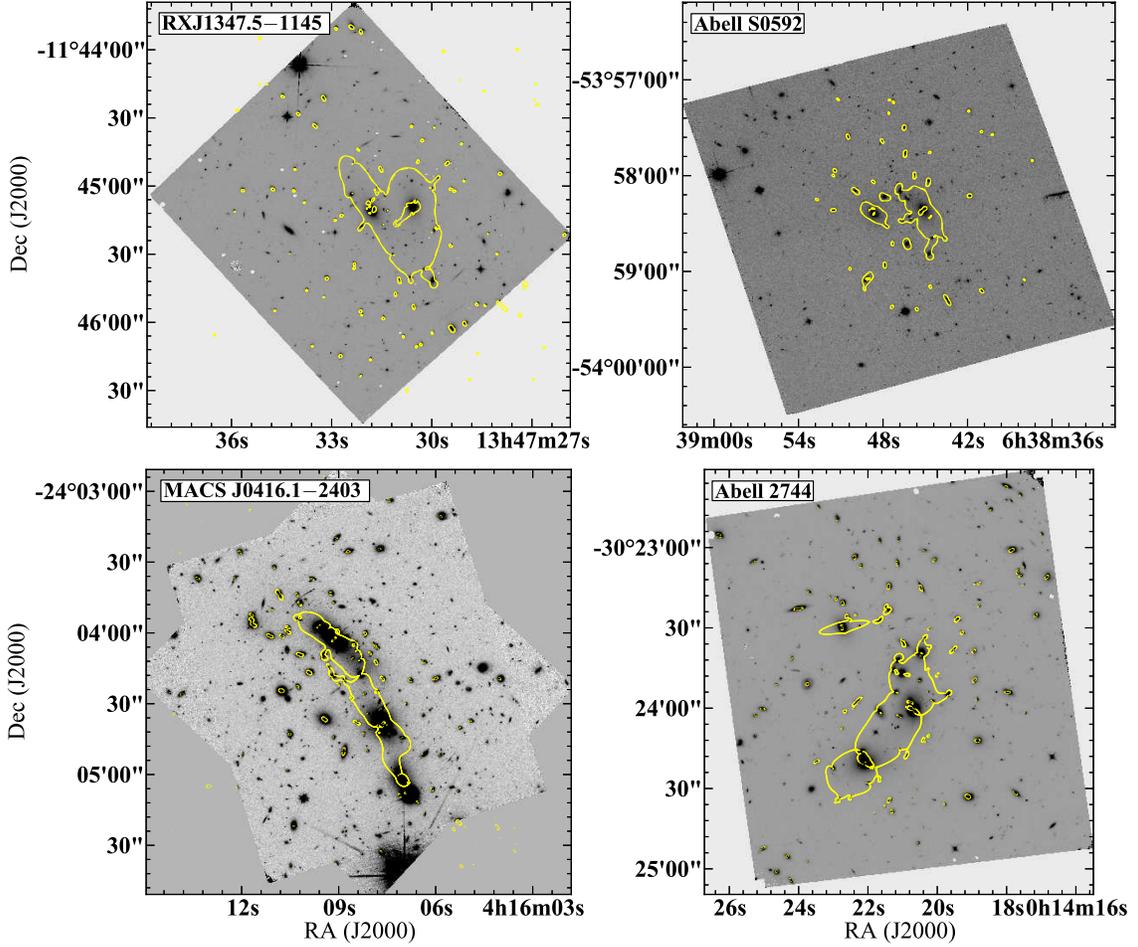}
\caption{Images of 4 lensing clusters obtained by {\it HST}. From the upper left panel to the lower right panel, the {\it HST}/WFC3 {\it F160W} image of RXJ1347.5$-$1145, the {\it HST}/ACS {\it F606W} image of Abell S0592, the {\it HST}/WFC3 {\it F160W} image of MACS J0416.1$-$2403, and the {\it HST}/WFC3 {\it F160W} image of Abell 2744, respectively. Yellow solid lines are critical lines at $z$ = 1.0 obtained by Kitayama et al.~in preparation, Oguri et al.~in preparation, and \cite{kawamata2016}. \label{fig:HST_image}}
\end{figure*}

In this paper, we present the results of a blind millimeter line emitter search using ALMA Band 6 data with only a single frequency tuning (i.e., observed frequency range $\simeq$ 8 GHz) toward four gravitational lensing clusters, RXJ1347.5$-$1145, Abell S0592, MACS J0416.1$-$2403, Abell 2744; images of these gravitational lensing clusters obtained by the {\it Hubble Space Telescope} ({\it HST}) are displayed in Figure \ref{fig:HST_image}. From our search, we constrain the CO luminosity functions at $z\lesssim1$ and the [CII] luminosity function at $z\simeq6$.

According to the predictions of semi-analytical models \citep[e.g.,][]{obreschkow2009a, obreschkow2009b, lagos2012, popping2016}, the number density of CO line emitters (i.e., CO luminosity function) evolve significantly at $z\lesssim1$, which is in marked contrast to the weak evolution at $z=1$--$4$ \citep{popping2016}. Because of the magnification effect of gravitational lensing clusters, we can constrain the fainter end of the CO luminosity function, which is difficult to observe previous unlensed blank field observations. Constraining the faint-end of CO luminosity functions ($L^\prime_\mathrm{CO}\lesssim$ 10$^{9}$ K km s$^{-1}$ pc$^2$) is particularly important because it is dominated by non-starburst galaxies, which are the main contributors to the cosmic SFRD. The faint-end of the CO luminosity functions are also affected by the CO spectral line energy distributions of galaxies, which reflect the density and temperature of the interstellar medium \citep[e.g., ][]{lagos2012, popping2016}. Furthermore, the [CII] luminosity function can be a useful tool to estimate the cosmic star formation rate density at $z\simeq$ 6, where the contribution from dusty galaxies to the cosmic SFRD is still uncertain.

This paper is structured as follows. Section \ref{sec:data_methods} presents the ALMA data and methods of our line emitter search. In Section \ref{sec:results}, we report results of our line emitter search. Then, we discuss the CO and [CII] luminosity functions in Section \ref{sec:discussion}. Section \ref{sec:summary} presents the summary and conclusion. Throughout this paper, we assume a $\Lambda$ cold dark matter cosmology with $\Omega_M = 0.3$, $\Omega_\Lambda = 0.7$, and $H_0 =$ 70 km s$^{-1}$ Mpc$^{-1}$.

\section{Data and methods} \label{sec:data_methods}

\subsection{ALMA data} \label{subsec:data}

Here, we present the ALMA data. Our ALMA Band 6 continuum observations were carried out as an ALMA Cycle 2 program (Project ID: 2013.1.00724.S, PI: H. Ezawa) on April 9 and 10, 2015, toward two gravitational lensing clusters (RXJ1347.5$-$1145 and Abell S0592). For the ALMA observation, 35--38 antennas were employed. The minimum and maximum baselines were 15.1 and 348.5 m, respectively. For RXJ1347.5$-$1145 (for Abell S0592), the phase calibrator was J1337$-$1257 (J0608$-$5456), the bandpass calibrators were J1337$-$1257 and J1256$-$0547 (J1107$-$4449 and J1058$-$0133), and the flux calibrators were Titan and Ganymede (Ganymede). The observed area, observed frequency, frequency setting, achieved continuum sensitivities, and synthesized beams are summarized in Table \ref{tab:cluster}.

We used additional Band 6 continuum observations toward another two gravitational lensing clusters (MACS J0416.1$-$2403 and Abell 2744) to expand our survey volume. These observations were also carried out as an ALMA Cycle 2 program (Project ID: 2013.1.00999.S, PI: F. Bauer). All data sets are public in the ALMA science archive. In Table \ref{tab:cluster}, we summarize the results of these continuum observations.

\begin{figure*}
\plotone{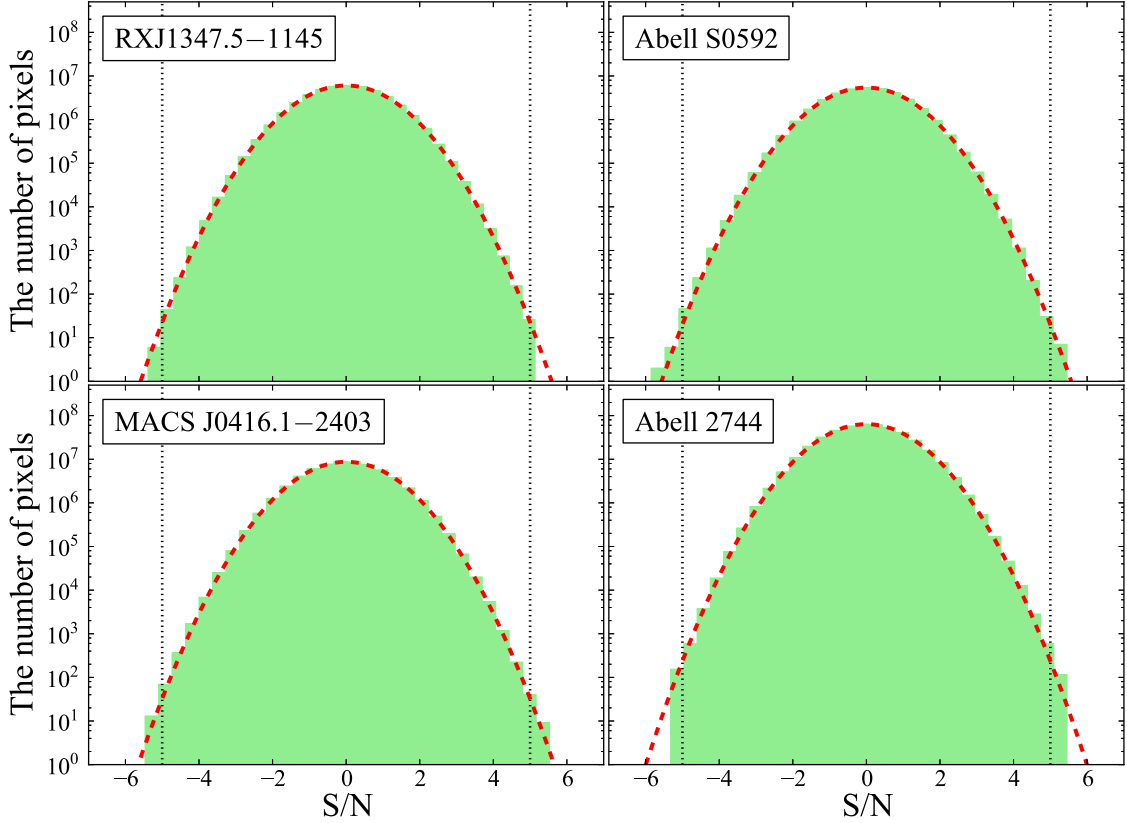}
\caption{The S/N distribution of 3D data cubes with 60 MHz binning. The red dashed line shows a Gaussian function. Black dotted lines indicate S/N = $\pm$ 5.  \label{fig:histo}}
\end{figure*}

From the calibrated measurement sets of clusters, we create three-dimensional (3D) data cubes against each spectral window with two different frequency resolutions, i.e., 60 and 100 MHz (corresponding to about 66.7 km s$^{-1}$ and 111 km s$^{-1}$ at 270 GHz, respectively). The calibrated visibilities are Fourier transformed using the task CLEAN in the Common Astronomy Software Application \citep[\texttt{CASA};][]{mcmullin2007}. In this study, we use cubes without beam deconvolution employing the CLEAN algorithm \citep{hogdom1974}, i.e., ``dirty cubes'', to search for line emitters, because no strong emission above 6$\sigma$ is found in these cubes (see Figure \ref{fig:histo}). The achieved angular resolutions of the 3D data cubes are approximately 1\arcsec--1.\arcsec5. Note that the frequency resolutions of the original data are about 35 km s$^{-1}$, 35 km s$^{-1}$, 18 km s$^{-1}$ and 18 km s$^{-1}$ for RXJ1347.5$-$1145, Abell S0592, MACS J0416.1$-$2403, and Abell 2744, respectively.

\begin{deluxetable*}{cccccccccc}[!]
\tabletypesize{\footnotesize}
\tablecaption{Our targets \label{tab:cluster}}
\tablecolumns{11}
\tablewidth{0pt}
\tablehead{
\colhead{Target} &
\colhead{$z_\mathrm{cluster}$} &
\colhead{$A$} &
\colhead{$\nu_\mathrm{obs}$} &
\colhead{$\sigma_\mathrm{cont.}$} &
\colhead{Synthesized beam} &
\colhead{$\Delta\nu$} &
\colhead{$\sigma_\mathrm{60\ MHz}$} &
\colhead{$\sigma_\mathrm{100\ MHz}$} & 
\colhead{$t_\mathrm{obs.}$} \\
\colhead{} & \colhead{} & \colhead{[arcmin$^2$]} & \colhead{[GHz]} &
\colhead{[$\mu$Jy beam$^{-1}$]} & \colhead{} & \colhead{[GHz]} & 
\colhead{[mJy beam$^{-1}$]} & \colhead{[mJy beam$^{-1}$]} & \colhead{[hrs.]} \\
\colhead{(1)} & \colhead{(2)} & \colhead{(3)} & \colhead{(4)} & \colhead{(5)} & 
\colhead{(6)} & \colhead{(7)} & \colhead{(8)} & \colhead{(9)} & \colhead{(10)} 
}
\startdata
\multirow{2}{*}{RXJ1347.5$-$1145}    & \multirow{2}{*}{0.451} & \multirow{2}{*}{4.75} & \multirow{2}{*}{265} & \multirow{2}{*}{155}  & \multirow{2}{*}{$1.\arcsec3\times0.\arcsec72\ (78^{\circ})$} & 255--259 & \multirow{2}{*}{1.4}  & \multirow{2}{*}{1.2}  & \multirow{2}{*}{2.26} \\
                    &       &     &      &                    &                         & 271--275 &      &    \\
\hline
\multirow{2}{*}{Abell S0592}         & \multirow{2}{*}{0.222} & \multirow{2}{*}{3.63} & \multirow{2}{*}{265} & \multirow{2}{*}{150}  & \multirow{2}{*}{$1.\arcsec2\times0.\arcsec75\ (87^{\circ})$} & 255--259 & \multirow{2}{*}{1.2}  & \multirow{2}{*}{1.0}  & \multirow{2}{*}{1.91} \\
                    &       &     &      &                    &                         & 271--275 &      &    \\
\hline
\multirow{2}{*}{MACS J0416.1$-$2403} & \multirow{2}{*}{0.397} & \multirow{2}{*}{4.45} & \multirow{2}{*}{263} & \multirow{2}{*}{73}   & \multirow{2}{*}{$1.\arcsec5\times0.\arcsec85\ (-84^{\circ})$} & 254--257  & \multirow{2}{*}{0.73} & \multirow{2}{*}{0.56} & \multirow{2}{*}{8.50} \\
                    &       &     &      &                      &                       & 269--272 &      &    \\
\hline
\multirow{2}{*}{Abell 2744}          & \multirow{2}{*}{0.308} & \multirow{2}{*}{4.26} & \multirow{2}{*}{263} & \multirow{2}{*}{91}   & \multirow{2}{*}{$1.\arcsec5\times1.\arcsec2\ (88^{\circ})$}  & 254--257 & \multirow{2}{*}{0.95} & \multirow{2}{*}{0.77} & \multirow{2}{*}{7.89} \\
                    &       &     &      &                    &                         & 269--272 &      &    \\
\enddata
\tablecomments{(1) Cluster name. (2) Redshifts of lensing clusters. {(3) Observed area.} (4) Central frequencies of observations. (5) Typical sensitivities of continuum maps. (6) Synthesized beam size of continuum map. Position angles of synthesized beams are given in parenthesis. {(7) Observed frequency setting.} (8) Typical sensitivities of 3D data cube with 60 MHz binning. (9) Typical sensitivities of 3D data cube with 100 MHz binning. (10) Total observation time.}
\end{deluxetable*}


\subsection{Methods of line emitter search} \label{subsec:methods}
First, we calculate the standard deviations in each channel and examine 3D signal-to-noise ratio cubes (S/N cubes) by dividing each data cube channel with its standard deviation. Note that we use the data cubes before the correction of the primary beam attenuation pattern to calculate the standard deviations. In Figure \ref{fig:histo}, we present the S/N distributions of the 3D data cube with 60 MHz binning. Next, we apply \texttt{CLUMPFIND} software \citep{williams1994} to the 3D S/N cubes to search for line emitter candidates with a peak S/N $>$ 5. We adopted the following parameters of CLUMPFIND: $\Delta S = 2\sigma$ and $S_\mathrm{start} = 4\sigma$, where $\Delta S$ is the contouring interval and $S_\mathrm{start}$ is the starting contour level as discussed in \citet{williams1994}. Finally, we remove spurious detections by eye; specifically, we deem line emitter candidates that were not detected with a S/N $>$ 3 in any channel adjacent to their peak channel as spurious and exclude them.

\begin{figure*}[!]
\epsscale{1.}
\plotone{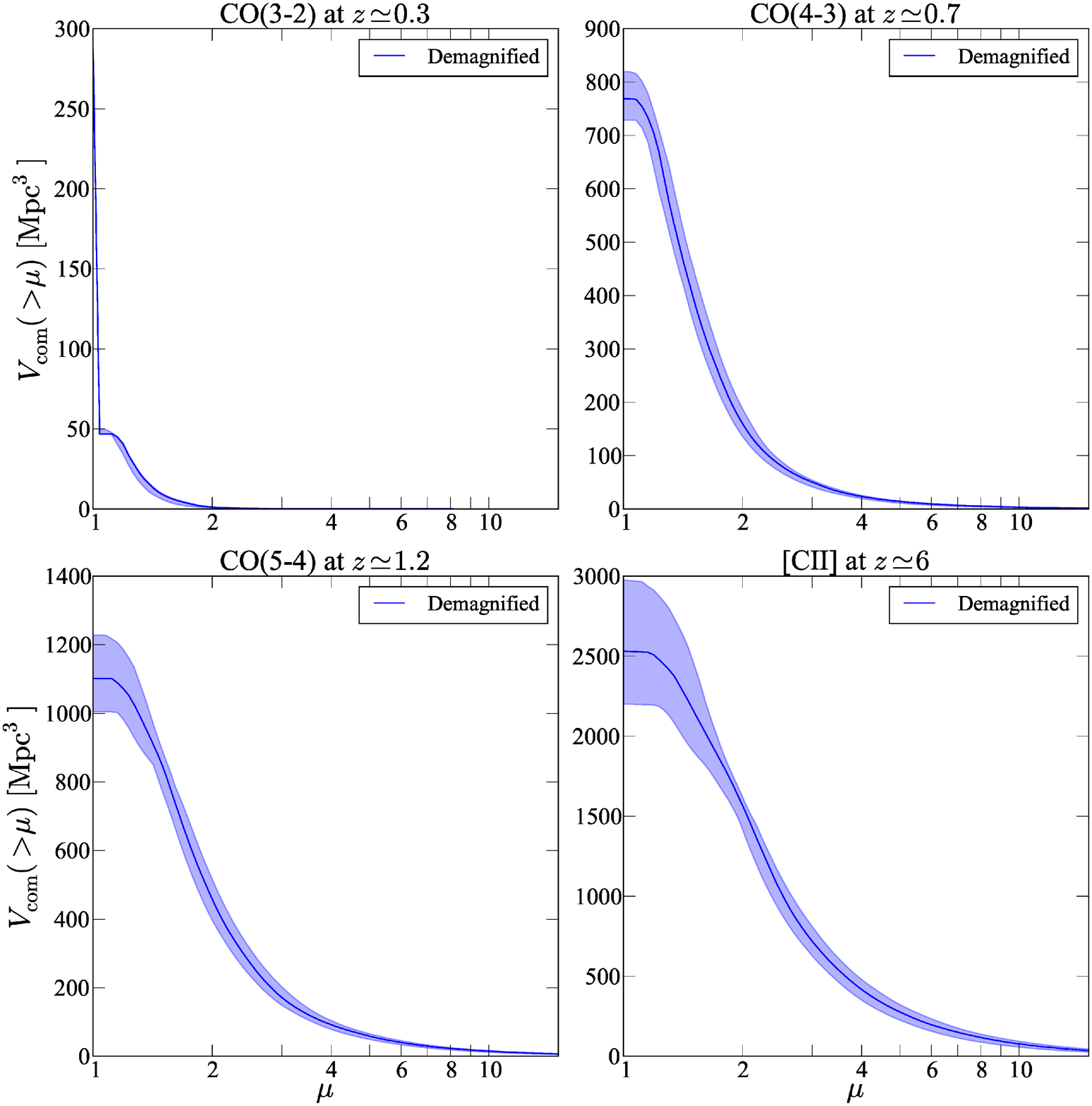}
\caption{The effective (i.e.~real) co-moving survey volume as a function of magnification factors ($\mu$). Blue shaded regions indicate model uncertainties (see section \ref{sec:discussion} for details).} \label{fig:mu_volume}
\end{figure*}

\begin{figure*}[!]
\epsscale{1.}
\plotone{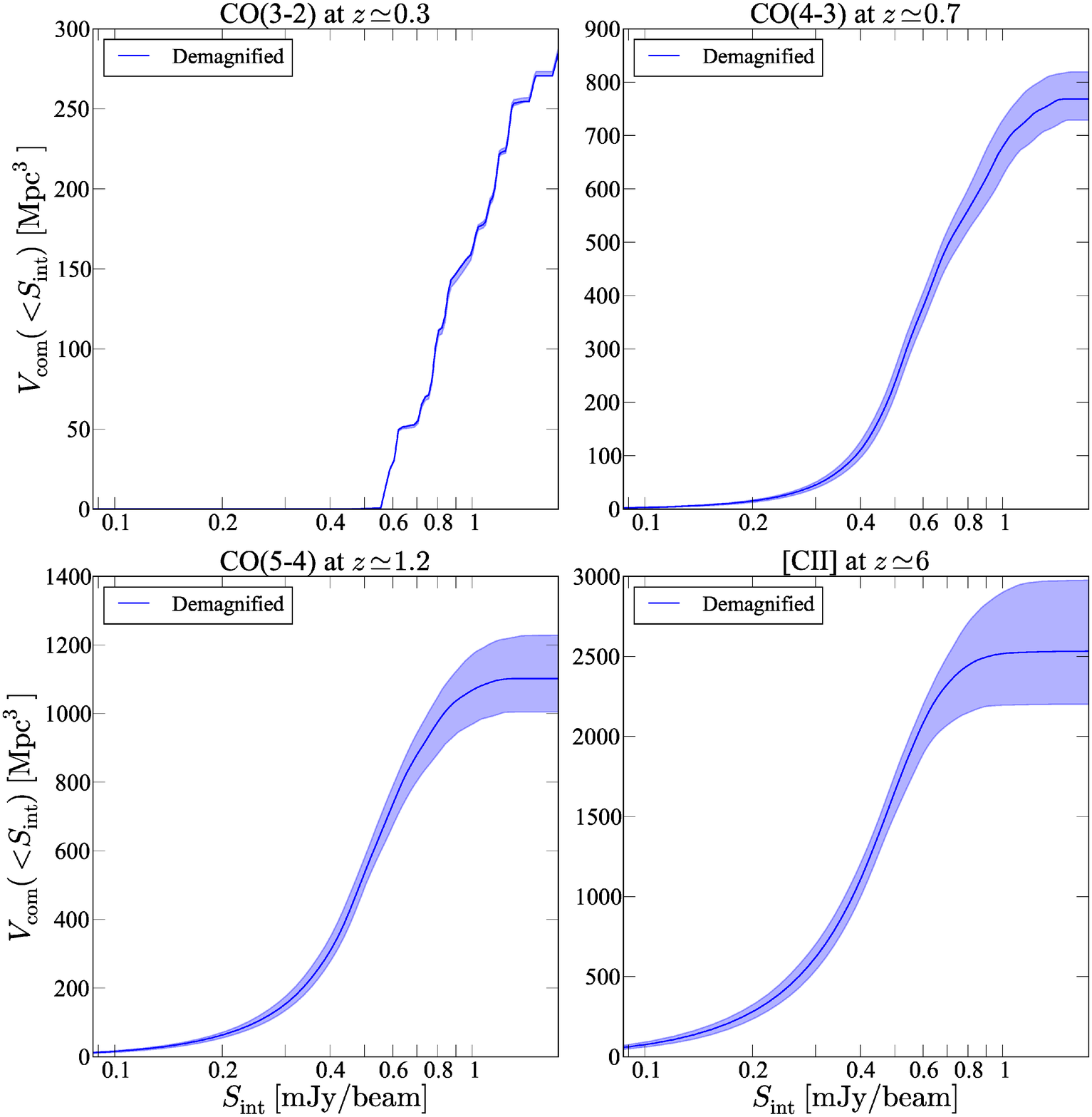}
\caption{The effective (i.e.~real) co-moving survey volume as a function of intrinsic 1$\sigma$ sensitivities. Blue shaded regions indicate model uncertainties (see section \ref{sec:discussion} for details).} \label{fig:volume}
\end{figure*}

\begin{deluxetable*}{cccccc}[!]
\tabletypesize{\small}
\tablecaption{The constraint on densities of line emitters \label{tab:CO_LF}}
\tablecolumns{6}
\tablewidth{0pt}
\tablehead{
\colhead{Line} &
\colhead{Redshift range} &
\colhead{$\log L^{\prime}_\mathrm{line}$} &
\colhead{$V_\mathrm{com}$} &
\colhead{Density}  \\
\colhead{} & \colhead{} & \colhead{\footnotesize{[K km s$^{-1}$ pc$^{2}$]}} & \colhead{\footnotesize{[Mpc$^3$]}} & \colhead{\footnotesize{[Mpc$^{-3}$]}}\\
\colhead{(1)} & \colhead{(2)} & \colhead{(3)} & \colhead{(4)} & \colhead{(5)} 
}
\startdata
\multirow{2}{*}{CO(3--2)} & 0.257--0.276, 0.335--0.357\tablenotemark{a} & 8.3--8.8  & 0.6778$^{+0.1666}_{-0.2799}$ & $<2.7^{+1.9}_{-0.5}$ \\ 
		                  & 0.286--0.271, 0.346--0.361\tablenotemark{b} & 8.8--9.3  & 285.5$^{+2.6}_{-0.4}$  & $<(6.4^{+0.0}_{-0.1}) \times 10^{-3}$ \\
\hline
\multirow{3}{*}{CO(4--3)} & \multirow{2}{*}{0.677--0.701, 0.780--0.808\tablenotemark{a}} & 8.3--8.8  & 10.94$^{+1.16}_{-1.99}$  & $<(1.7^{+0.4}_{-0.2}) \times 10^{-1}$\\
                          & \multirow{2}{*}{0.695--0.714, 0.794--0.815\tablenotemark{b}} & 8.8--9.3  & 316.8$^{+26.5}_{-27.0}$  & $<(5.8^{+0.5}_{-0.5}) \times 10^{-3}$ \\
                          &                                                              & 9.3--9.8  & 768.5$^{+51.0}_{-39.9}$  & $<(2.4^{+0.1}_{-0.1}) \times 10^{-3}$ \\
\hline
\multirow{3}{*}{CO(5--4)} & \multirow{2}{*}{1.10--1.13, 1.22--1.26\tablenotemark{a}} & 8.5--9.0  & 47.42$^{+5.35}_{-6.88}$ & $<(3.9^{+0.7}_{-0.4}) \times 10^{-2}$\\
                          & \multirow{2}{*}{1.12--1.14, 1.24--1.27\tablenotemark{b}} & 9.0--9.5  & 642.5$^{+50.8}_{-50.8}$ & $<(2.9^{+0.2}_{-0.2}) \times 10^{-3}$ \\
                          &                                                          & 9.5--10.0 & 1102$^{+126}_{-96}$  & $<(1.7^{+0.2}_{-0.2}) \times 10^{-3}$ \\
\hline
\multirow{3}{*}{[CII] 158 $\mu$m} & \multirow{2}{*}{5.91--6.01, 6.34--6.45\tablenotemark{a}} & 8.2--8.7\tablenotemark{c}  & 216.0$^{+34.9}_{-98.5}$ & $<(8.5^{+1.8}_{-1.1}) \times 10^{-3}$\\
                                  & \multirow{2}{*}{5.99--6.07, 6.40--6.48\tablenotemark{b}} & 8.7--9.2\tablenotemark{c}  & 1896$^{+106}_{-178}$  & $<(9.7^{+1.0}_{-0.5}) \times 10^{-4}$ \\
                                  &                                                          & 9.2--9.7\tablenotemark{c}  & 2532$^{+444}_{-331}$  & $<(7.3^{+1.1}_{-1.1}) \times 10^{-4}$ \\
\enddata
\tablecomments{(1) Observed line. (2) Observed redshift range. (3) Intrinsic (i.e., demagnified) line luminosities. (4) Co-moving survey volume. (5) The 1$\sigma$ confidence upper limits on the densities of line emitters, which are calculated by using the Poisson statistics by \citet{gehrels1986}.}
\tablenotetext{a}{Observed redshift range of RXJ1347.5$-$1145 and Abell S0592.}
\tablenotetext{b}{Observed redshift range of MACS J0416.1$-$2403 and Abell 2744.}
\tablenotetext{c}{For [CII] 158 $\mu$m line, units of line luminosities are $L_\odot$.}
\end{deluxetable*}

\section{Results} \label{sec:results}
We do not detect any significant line-emission in our search. The S/N distributions are well fitted by Gaussian functions, which also support non-detections (Figure \ref{fig:histo}). The typical apparent 1$\sigma$ noise levels of the data cubes are $\sigma\simeq$ 1.4, 1.2, 0.73, and 0.95 mJy beam$^{-1}$ with 60 MHz binning and $\sigma\simeq$ 1.2, 1.0, 0.56, and 0.77 mJy beam$^{-1}$ with 100 MHz binning for RXJ1347.5$-$1145, Abell S0592, MACS J0416.1$-$2403, and Abell 2744, respectively (Table \ref{tab:cluster}). Thus, if we assume $\Delta V=$ 200 km s$^{-1}$, as presumed in \citet{decarli2016}, the 3$\sigma$ limiting apparent CO luminosities are estimated to be $\mu L^{\prime}_{\mathrm{CO}}$ $\simeq$ $5.5\times10^8$, $1.8\times10^{9}$, and $2.9\times10^{9}$ K km s$^{-1}$ pc$^2$ for CO(3--2) at $z\simeq0.3$, CO(4--3) at $z\simeq0.7$, and CO(5--4) at $z\simeq1.2$, respectively. Note that $\Delta V$ and $\mu$ are the velocity-width and the gravitational lensing magnification factor, respectively. For the [CII] 158 $\mu$m line at $z\simeq$ 6, the 3$\sigma$ limiting apparent [CII] luminosities are estimated to be $\mu L_\mathrm{[CII]}\simeq$ $1.0\times10^{9}$, $8.5\times10^{8}$, $4.7\times10^{8}$, $6.5\times10^{8}$ $L_\odot$. In the case of the [CII] 158 $\mu$m line, we assume $\Delta V$ = 300 km s$^{-1}$ as explained in \citet{aravena2016}.

If we adopt a detection threshold of S/N = 4.0, there is a tentative detection of a line emitter at ($\alpha_\mathrm{J2000}$, $\delta_\mathrm{J2000}$) = (13$^\mathrm{h}$47$^\mathrm{m}$30$^\mathrm{s}$.13, $-$11$^\circ$45$^\prime$26$^{\prime\prime}$.59) in RXJ1347.5$-$1145 (see Figures in Appendix; hereafter RXJ1347-emitter1). RXJ1347-emitter1 is detected with 4.5$\sigma$ at the peak channel in the 60 MHz-binning data cube and detected with 4.3$\sigma$ at next to the peak channel. RXJ1347-emitter1 is also detected with 5.8$\sigma$ in the 100 MHz-binning data cube, but only detected at the peak channel. RXJ1347-emitter1 has no optical/NIR counterpart (see Figures in Appendix). RXJ1347-emitter1 is not detected in ALMA continuum map. However, the negative tail of the noise distribution of the 60 MHz-binning data cubes also extends to S/N = $-4.5$ (see Figure \ref{fig:histo}) and is only detected at the peak channel in the 100 MHz-binning data cube. Thus, we treat RXJ1347-emitter1 as the ``line emitter candidate'' in this paper. Further details of RXJ1347-emitter1 will be provided in Appendix.

\citet{gonzaez2017} also search for line emitters using MACS J0416.1$-$2403 and Abell 2744 data, and report some detections (6 in MACS J0416.1$-$2403, 3 in Abell 2744). This discrepancy is simply because our criterion are more conservative than their criterion.

\section{CO and [CII] luminosity functions} \label{sec:discussion}

We define luminosity bins to range from our 3$\sigma$ limiting apparent luminosity (see Section \ref{sec:results}) to a 0.5-dex higher luminosity. Because of the magnification due to gravitational lensing, we can search for lower line luminosities than the 3$\sigma$ limiting apparent line luminosities. Accordingly, we adopt three intrinsic (i.e., demagnified) luminosity bins as displayed in Table \ref{tab:CO_LF}. Note that for CO(3--2), we define two intrinsic luminosity bins, because the survey volume for the lowest intrinsic luminosity bin becomes very small as explained later in this section.

To constrain the CO and [CII] luminosity functions, it is necessary to estimate the co-moving survey volume. For this purpose, we used gravitational lensing models constructed with the \texttt{GLAFIC} software, which adopt a standard $\chi^2$ minimization to determine the best-fit mass model \citep[see][for details]{oguri2010}. For MACS J0416.1$-$2403 and Abell 2744 we use public \texttt{GLAFIC} mass models (version 3.0) that are available at Space Telescope Science Institute (STScI) website\footnote{https://archive.stsci.edu/prepds/frontier/lensmodels/} \citep{kawamata2016}. For the other two clusters, we use mass models obtained by Kitayama et al.~in preparation (for RXJ1347.5$-$1145) and Oguri et al.~in preparation (for Abell S0592). These models are constructed in a similar way to \cite{kawamata2016}.

In Figure \ref{fig:mu_volume} and Figure \ref{fig:volume}, we plot the effective (i.e.~real) co-moving survey volume as a function of magnification factors and intrinsic 1$\sigma$ sensitivities, respectively. For CO(3--2), the demagnified survey volume is small, especially in the high magnification area. This is because the CO(3--2) emitters at $z\sim0.3$ are located in front of the gravitational lensing clusters at $z\lesssim 0.3$ and are thus not affected by gravitational lensing. This means that the non-null contribution at $\mu>$ 1 values only comes from Abell S0592.

We use Markov Chain Monte Carlo (MCMC) methods to estimate model uncertainties as with the case of \cite{kawamata2016}. For MACS J0416.1$-$2403 and Abell 2744, results of MCMC methods are also available at STScI website$^1$. In MCMC methods, we change following parameters; virial mass, positions, ellipticity, position angle, concentration parameters, velocity dispersion, truncation radius, dimension less parameter $\eta$, and redshifts of lensed galaxies \citep[see][for details]{kawamata2016}. The resulting MCMC chain typically consists of hundreds of thousands of points. From the MCMC chain we randomly pick 100 parameter sets to estimate the error in our volume estimate from the mass model uncertainty. Specifically, we estimate the co-moving volume for each parameter set, repeat it for the 100 parameter set, and derive the model uncertainties. Here, we define the range between the maximum and minimum co-moving survey volume as the model uncertainty. Our co-moving survey volume of each luminosity bin and the 1$\sigma$ confidence upper limits on the densities of line emitters, which are calculated by using the Poisson statistics by \citet{gehrels1986} are summarized in Table \ref{tab:CO_LF}. Note that our estimated errors can be underestimated because we do not include systematic errors between different lens models. For example, the area with magnification between 5 and 10 in Abell 2744 can change by almost $\sim$ 20\% between different lens models released in STScI website$^1$ \citep{wang2015, kawamata2016, priewe2017}.

Note that the limiting luminosities do not depend on the assumed line profiles. For instance, if we adopt the limiting luminosities following our detection criterion explained in Section \ref{subsec:methods} (i.e., it is detected with 5$\sigma$ flux density in one channel, and 3$\sigma$ in a neighboring channel), results do not change. For example, for CO(4--3), the apparent limiting luminosity is estimated to be $\mu L^{\prime}_{\mathrm{CO(4\mathchar`-3)}}$ $\simeq$ $10^{9.2}$ K km s$^{-1}$ pc$^2$ (in 60 MHz-binning cubes), which is comparable with the limiting luminosity presented in Section \ref{sec:results}. In this case, the typical magnification factor corresponding to the faintest luminosity bin in the Table \ref{tab:CO_LF} (i.e., $L^{\prime}_{\mathrm{CO(4\mathchar`-3)}}$ $\simeq$ $10^{8.5}$ K km s$^{-1}$ pc$^2$) is $\mu\simeq$ 5. According to Figure \ref{fig:mu_volume}, the co-moving survey volume corresponding to this case (i.e., $\mu\simeq$ 5) is estimated to be $V_\mathrm{com}\simeq$ 14 Mpc$^3$, which is comparable with Table \ref{tab:CO_LF}.

\begin{figure*}[!]
\epsscale{1.2}
\plotone{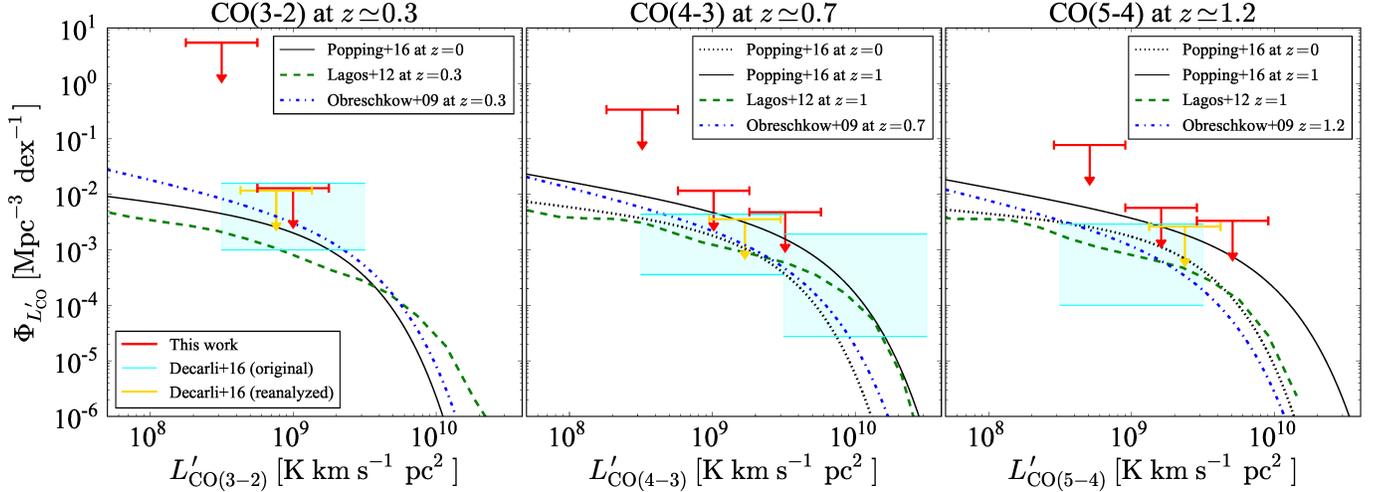}
\caption{Comparison of our blind line emitter search with the empirically derived CO luminosity functions and previous observations. Our results are presented as red symbols. Black solid lines, green dashed lines, and blue dot-dashed lines are the empirically derived CO luminosity functions from \citet{obreschkow2009a, obreschkow2009b}, \citet{lagos2012}, and \citet{popping2016}, respectively. Cyan shaded regions are the results of the ASPECS \citep{decarli2016}. Yellow symbols are the results of our reanalysis of the ASPECS data (see Section \ref{sec:CO_LF}).
\label{fig:CO_LF}}
\end{figure*}

\subsection{CO luminosity functions} \label{sec:CO_LF}

In Figure \ref{fig:CO_LF}, we show our constraints on the CO luminosity functions (red symbols). In order to avoid CO excitation uncertainties, we do not convert $L^{\prime}_{\mathrm{CO(3\mathchar`-2)}}$, $L^{\prime}_{\mathrm{CO(4\mathchar`-3)}}$, and $L^{\prime}_{\mathrm{CO(5\mathchar`-4)}}$ into $L^{\prime}_{\mathrm{CO(1\mathchar`-0)}}$ in this paper. In gravitational lensing clusters, the effective survey volumes with a large magnification factor is small as shown in Figure \ref{fig:mu_volume} and Figure \ref{fig:volume}. This is the reason why our constraints on the CO luminosity functions at the faintest intrinsic luminosity bins are not strong. We only plot the best-fitting case in Figure \ref{fig:CO_LF}, because model uncertainties on luminosity functions are small (see Table {\ref{tab:CO_LF}}).

In the same plot (Figure \ref{fig:CO_LF}), we also show the predictions based on semi-analytical cosmological models by \citet{obreschkow2009a, obreschkow2009b}, \citet{lagos2012}, and \citet{popping2016}. As shown in Figure \ref{fig:CO_LF}, our constraints are consistent with their predictions. 

We also plot the latest results of the ALMA SPECtroscopic Survey in the \textit{Hubble} Ultra-Deep Field \citep[ASPECS][cyan shaded regions]{walter2016, decarli2016}. However, they only use peak values to identify line emitters. To make a fair comparison, we reanalyze their ALMA Band 6 data (Project ID: 2013.1.00718.S, PI: M.~Aravena) following our procedure, which is explained in Section \ref{subsec:methods}. In our procedure, we detect two emission lines, which are detected in ASPECS as 1mm.1 and 1mm.2 \citep{walter2016, decarli2016}. According to \citet{decarli2016}, these two lines represent the CO emission from one line emitter at $z=2.54$. Therefore, we can only place upper limits on the CO luminosity functions at $z\lesssim$ 1 (yellow symbols in Figure \ref{fig:CO_LF}) from the ASPECS data. As shown in Figure \ref{fig:CO_LF}, our constraints on the CO luminosity functions are consistent with the ASPECS results at similar luminosity ranges ($L^\prime_\mathrm{CO}\sim$ 10$^{9}$ K km s$^{-1}$ pc$^2$). Although the upper limit is about 1--2 orders of magnitude larger than the predictions of semi-analytical models, we can expand the range of luminosity to $\gtrsim$ 0.5-dex lower than previous observations, although the current constraints are very coarse.

Based on our upper limits, we constrain the density evolution of the CO luminosity functions between $z=0$ and $z\simeq1$. As shown in Figure \ref{fig:CO_LF}, the evolution of the CO luminosity functions between $z=0$ and $z\simeq1$ are consistent with the predictions of semi-analytical models by \citet{obreschkow2009a, obreschkow2009b}, \citet{lagos2012} and \citet{popping2016}, although the constraints to date are not stringent yet.

\begin{figure}[!]
\epsscale{1.2}
\plotone{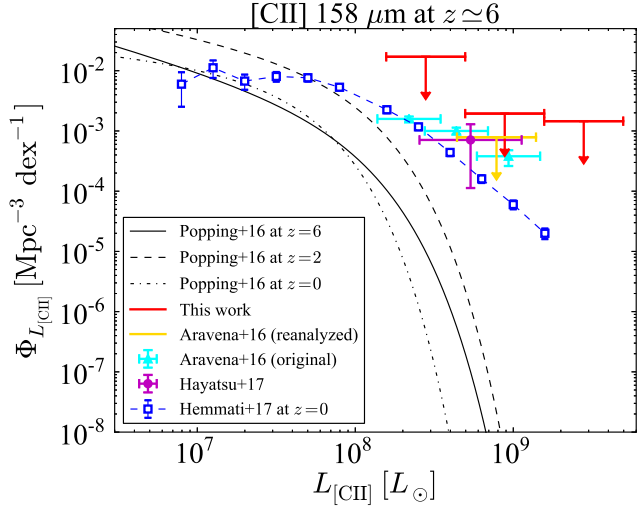}
\caption{Comparison of our blind line emitter search with the empirically derived [CII] luminosity functions and previous observations. Our results are presented as red symbols. The black solid line, the dashed line and the dot-dashed line is empirically derived [CII] luminosity functions at $z=6$, $z=2$, and $z=0$ from \citet{popping2016}, respectively. The cyan triangle and the magenta circle indicates the observational results of the ASPECS \citep{aravena2016} and Hayatsu et al.~(2017 submitted), respectively. The yellow symbol is result of our reanalysis of ASPECS data. Blue squares represent the observed [CII] luminosity function at $z=0$ \citep{hemmati2017}.
\label{fig:cii_LF}}
\end{figure}

\subsection{[CII] luminosity function} \label{sec:CII_LF}

We display our constraints on the [CII] luminosity function at $z\simeq$ 6 in Figure \ref{fig:cii_LF}. As with the case of CO luminosity functions, we only plot the best-fitting case. In the same plot, we also show the predictions based on semi-analytical cosmological models by \citet{popping2016} and observational results of ASPECS \citep{aravena2016} and \citet{hayatsu2017}. As with the case of the CO luminosity functions, we show the results of the ASPECS data reanalysis (see Sec. \ref{sec:CO_LF} for details). We also plot the observed [CII] luminosity function at $z=0$ observed by the {\it Herschel Space Observatory} \citep{hemmati2017}.

Although the upper limits are significantly higher than the prediction of \cite{popping2016}, our results are still consistent with previous observational results (Figure \ref{fig:cii_LF}). Indeed, recent observations suggest that the semi-analytical models underestimate the number density of [CII] emitters \citep[e.g.,][]{swinbank2014, aravena2016, miller2016, hemmati2017, hayatsu2017} at the luminosity range of $L_\mathrm{[CII]}\gtrsim$ 10$^8$ $L_\odot$. Thus, our results support previous observations at the luminosity range of $L_\mathrm{[CII]}\sim$ 10$^8$--10$^{10}$ $L_\odot$.

\citet{popping2016} predict that the [CII] luminosity function at $z=6$ returns to a level similar to that of $z=0$. This is also suggested by observational studies \citep[e.g., ][]{aravena2016, hemmati2017}, regardless of [CII] luminosity function shape. As shown in Figure \ref{fig:cii_LF}, our results are also consistent with the prediction.

\section{Summary \& conclusion}
\label{sec:summary}
We carried out a blind search for millimeter line emitters using ALMA band 6 data with a single frequency tuning toward four gravitational lensing clusters. We did not detect any line emitters with a peak S/N $>$ 5, although we did find one line emitter candidate (RXJ1347-emitter1) with a peak S/N = 4.5 in the 60 MHz-binning data cube. 

We placed upper limits on the CO(3--2), CO(4--3), and CO(5--4) luminosity functions at $z\simeq$ 0.3, 0.7, and 1.2, respectively. Because of the magnification effect of gravitational lensing clusters, the new data provide the first constraints on the CO and [CII] luminosity functions at unprecedentedly low luminosity levels, i.e., down to $\lesssim$ 10$^{-3}$--10$^{-1}$ Mpc$^{-3}$ at $L^\prime_\mathrm{CO}\sim$ 10$^8$--10$^{10}$ K km s$^{-1}$ pc$^2$. These results are consistent with the predictions of semi-analytical models. Our constraint is comparable with the latest results of the ALMA spectroscopic scan observation of ASPECS at similar luminosity ranges ($L^\prime_\mathrm{CO}\sim$ 10$^{9}$ K km s$^{-1}$ pc$^2$). However, we can expand the range of luminosity to $\gtrsim$ 0.5-dex lower than previous observations. Our constraint on the evolution of CO luminosity function between $z=0$ and $z\simeq1$ are consistent with the predictions of semi-analytical models by \citet{obreschkow2009a, obreschkow2009b}, \citet{lagos2012} and \citet{popping2016}, although the constraints to date are not stringent yet.

We also placed upper limits on the [CII] luminosity function at $z\simeq$ 6. Although the upper limits are significantly higher than the prediction of the semi-analytical model, our results are still consistent with previous observational results. Our results are consistent with the scenario that the [CII] luminosity function returns to a level similar to that of $z=0$ at $z\simeq6$.

The total observation time of our data is comparable with ASPECS ($\sim$ 20 hours at Band 6). Therefore, this study demonstrates that not only the spectroscopic scan observations, but also the wide observations with a single frequency tuning toward gravitational lensing clusters are useful for constraining the CO and [CII] luminosity functions. We will also be able to apply stronger constraints by adding more ALMA Cycle 3 or 4 data toward gravitational lensing clusters, which will become public soon.


\acknowledgments
We thank the referee for the comments, which improve the manuscript.
This paper makes use of the following ALMA data: ADS/JAO.ALMA\#2013.1.00724.S, 2013.1.00999.S, and 2013.1.00718.S.
ALMA is a partnership of ESO (representing its member states), NSF (USA), and NINS (Japan) together with NRC (Canada), NSC and ASIAA (Taiwan), and KASI (Republic of Korea) in cooperation with the Republic of Chile.
The Joint ALMA Observatory is operated by ESO, AUI/NRAO, and NAOJ.
Data analysis was partly carried out on the common-use data analysis computer system at the Astronomy Data Center (ADC) of the National Astronomical Observatory of Japan.
Y.~Yamaguchi is thankful for the JSPS fellowship.
H.~Umehata is supported by JSPS Grant-in-Aid for Research Activity Start-up (16H06713).
T.~Kitayama, K.~Kohno, and Y.~Matsuda acknowledge support from JSPS KAKENHI Grant Numbers 25400236, 25247019, and 17H04831, respectively.
This work was supported in part by World Premier International Research Center Initiative (WPI Initiative), MEXT, Japan, and JSPS KAKENHI Grant Number 26800093 and 15H05892.
This research made use of the NASA/IPAC Extragalactic Database (NED), which is operated by the Jet Propulsion Laboratory, California Institute of Technology, under contract with the National Aeronautics and Space Administration.




\appendix

\subsection{Line emitter candidate; ``RXJ1347-emitter1''}
\label{sec:emitter-cand}
\begin{figure}[h]
\epsscale{1.}
\plotone{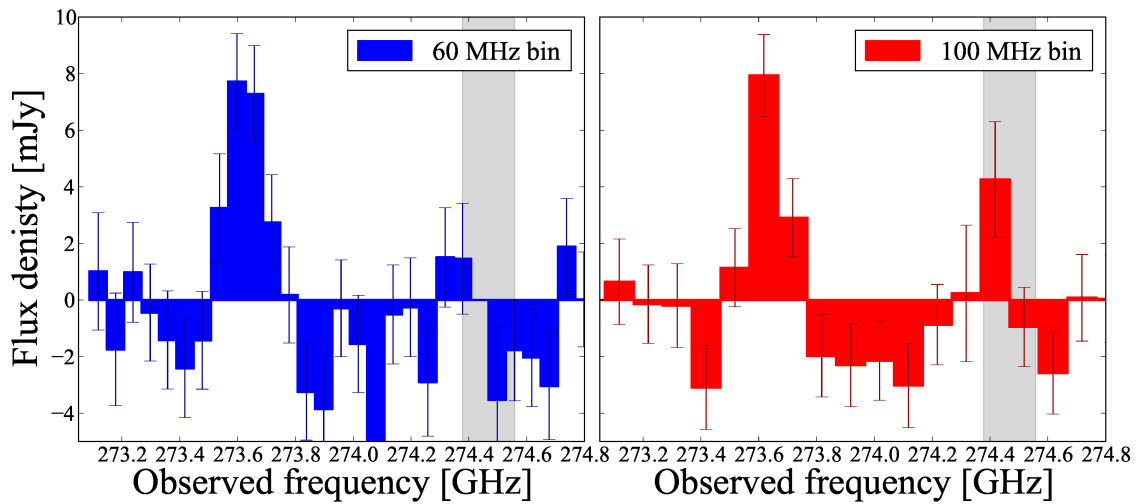}
\caption{From left to right, the spectrum of RXJ1347-emitter1 with 60 MHz binning and 100 MHz binning with 1$\sigma$ errorbars, respectively. The gray shaded regions indicates the frequency range of an atmospheric absorption line caused by ozone.
\label{fig:emitter_cand}}
\end{figure}

We find a line emitter candidate at ($\alpha_\mathrm{J2000}$, $\delta_\mathrm{J2000}$) = (13$^\mathrm{h}$47$^\mathrm{m}$30$^\mathrm{s}$.13, $-$11$^\circ$45$^\prime$26$^{\prime\prime}$.59). In Figure \ref{fig:emitter_cand}, we display the spectrum of RXJ1347-emitter1. RXJ1347-emitter1 is detected with S/N = 4.5 at the peak channel and with S/N = 4.3 at next to the peak channel in the 60-MHz-binning data. In the 100-MHz-binning data, it is detected with S/N = 5.8 at the peak channel. Although negative tail of the noise distribution of the 60-MHz-binning data extends to S/N = $-4.5$, we do not detect any pixels with S/N $<$ $-5.8$ in the 100-MHz-binning data. We have no atmospheric absorption lines around the peak frequency of RXJ1347-emitter1. There are no astronomical absorption features in the 3D data cubes of bandpass calibrators.

\begin{figure*}[h]
\epsscale{1.}
\plotone{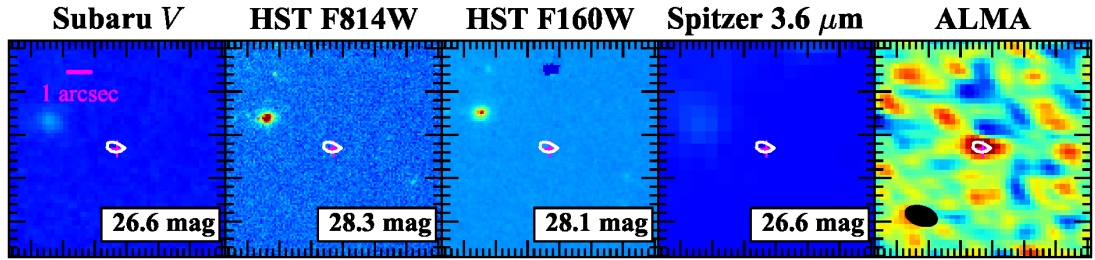}
\caption{From left to right: Subaru/Suprime-Cam {\it V} band, {\it HST}/ACS {\it F814W}, {\it HST}/WFC3 {\it F160W}, {\it Spitzer}/IRAC 3.6 $\mu$m, and ALMA velocity-integrated images of RXJ1347-emitter1, respectively. The magenta cross indicates the peak position of RXJ1347-emitter1. The white contour shows the 4$\sigma$ level of the ALMA velocity-integrated image. The black symbol is the synthesized beam of ALMA. Insert magnitudes are apparent 3$\sigma$ limiting magnitudes obtained by \citet{postman2012}, \citet{umetsu2014}, and \citet{huang2016}.
\label{fig:multi}}
\end{figure*}

As shown in Figure \ref{fig:multi}, RXJ1347-emitter1 has no counterpart at optical/NIR wavelengths. Thus, RXJ1347-emitter1 can be a [CII] 158 $\mu$m emitter at $z=$ 5.95 rather than a CO emitter at $z\sim$ 1, if it is a real line emitter. To confirm whether RXJ1347-emitter1 is a real detection or a spurious detection and determine redshift, future ALMA follow-up observation is needed.

\end{document}